\documentclass[12pt]{iopart}
\usepackage{graphicx}
\begin{document}

\title[]{Fluctuations and Correlations in STAR}

\author{Gary D. Westfall\dag\ for the STAR Collaboration}

\address{\dag\ National Superconducting Cyclotron Laboratory and Department of Physics and Astronomy, Michigan
State University, East Lansing, Michigan 48824-1321 USA\\
westfall@nscl.msu.edu}

\begin{abstract}
We report measurements for the balance function, $p_{t}$ fluctuations, and net charge fluctuations from Au+Au collisions at $\sqrt{s_{NN}}$ = 20, 130, and 200 GeV as well as p+p and d+Au collisions at $\sqrt{s_{NN}}$ = 200 GeV using STAR at RHIC.  For Au+Au collisions at 200 GeV, we observe a narrowing of the balance function in central collisions.  We observe dynamic $p_{t}$ fluctuations at all incident energies.  Observables related to $p_{t}$ fluctuations and net charge fluctuations are similar for peripheral Au+Au collisions and inclusive p+p collisions while central Au+Au collisions deviate significantly from HIJING predictions.
\end{abstract}




{\em Introduction.} The study of correlations and fluctuations can provide evidence for the production of the quark gluon plasma (QGP) in relativistic heavy ion collisions.[1-16]  Various theories predict that the production of a QGP phase in relativistic heavy ion collisions could produce significant event-by-event correlations and fluctuations in temperature, transverse momentum, multiplicity, and conserved quantities such as net charge.   Several recent experimental studies at the SPS[17-19] and at RHIC[20-24] have focused on the study of fluctuations in relativistic heavy ion collisions. In this paper we present studies of the balance function, $p_{t}$ fluctuations, and net charge fluctuations for p+p, d+Au, and Au+Au collisions at $\sqrt{s_{NN}}$ = 20, 130, and 200 GeV using STAR at RHIC.

{\em Balance Function.} The balance function\cite{balance_theory} may be sensitive to whether the transition to a hadronic phase was delayed, as expected if the quark-gluon phase were to persist for a substantial time.  In Ref. \cite{star_balance} we observed that the balance function narrows in central collisions of Au+Au at $\sqrt{s_{NN}}$ = 130 GeV, which is consistent with trends predicted by models incorporating delayed hadronization.  In contrast, HIJING calculations for the widths of the balance function show no centrality dependence.  Details of the balance function analysis can be found in Ref. \cite{star_balance}.

Here we present the balance function for p+p, d+Au, and Au+Au collisions at $\sqrt{s_{NN}}$ = 200 GeV.  The balance function for all charged particles for Au+Au collisions is shown in Fig. 1 for nine centrality bins as a function of the relative pseudorapidity, $\Delta\eta$.  Extracting the width of these balance functions using a weighted average for $0.2 \le \Delta\eta \le  2.0$, we observe that the width of the balance function for Au+Au collisions gets smaller in central collisions.  In Fig. 1 we show the width of the balance function as a function of the number of participating nucleons, $N_{part}$ for p+p, d+Au, and Au+Au collisions.  The balance function widths scale smoothly with $N_{part}$.  In contrast, widths predicted using HIJING calculations for Au+Au show little centrality dependence and are similar to those measured for p+p collisions.

\begin{figure}
\includegraphics[width=15.7cm]{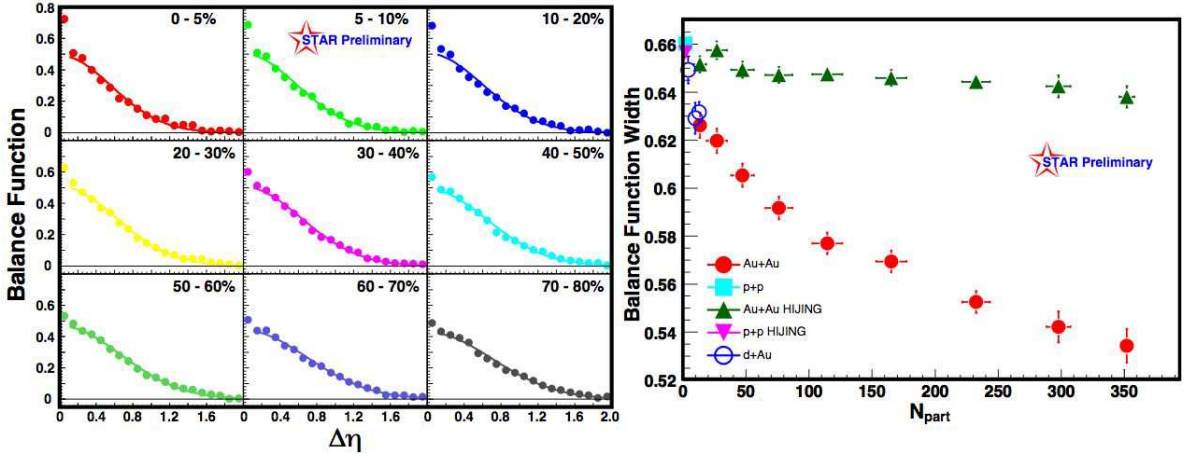}
\caption{\label{fig:fig1} The left panel shows the balance function versus $\Delta\eta$ for all charged particle pairs from Au+Au collisions at $\sqrt{s_{NN}}$ = 200 GeV as a function of centrality.  Lines are drawn to guide the eye.  The right panel shows the balance function widths for p+p, d+Au, and Au+Au collisions at $\sqrt{s_{NN}}$ = 200 GeV as a function the number of participating nucleons.  Also shown are HIJING calculations for p+p and Au+Au.}
\end{figure}

{\em $p_{t}$ Fluctuations.} In Fig.2 we show histograms of the average transverse momentum per event, $\langle p_{t} \rangle$, for Au+Au collisions at $\sqrt{s_{NN}}$ = 20, 130, and 200 GeV for all charged particles with $0.1 GeV/c < p_{t} < 2.0 GeV/c$ and $|\eta| < 1.0$ for the 5\% most central events at each beam energy.  Histograms for real data and mixed events are shown for all three energies. The histograms are fit with gamma distributions shown by the solid lines. The distributions for real events are wider than those for mixed events demonstrating that non-statistical fluctuations exist at all three energies. To isolate the dynamical fluctuations and minimize contributions from trivial statistical effects, we employ the two particle correlation $\langle \Delta p_{t,i} \Delta p_{t,j} \rangle$\cite{fluc_collective}.  In Fig. 2 we show the quantity $\sqrt {\left\langle {\Delta p_{t,i}\Delta p_{t,j}} \right\rangle }/\left\langle {\left\langle {p_t} \right\rangle } \right\rangle$ as a function of centrality for Au+Au collisions at 20, 130, and 200 GeV, where$ \left\langle {\left\langle {p_t} \right\rangle } \right\rangle$ is the inclusive average transverse momentum for each system and centrality bin.  These results are not corrected for short range correlations or resonance decay.  We observe that this quantity, which should be proportional to the correlation per particle, decreases with centrality but shows little dependence on the incident energy.  Results for the same quantity from Pb+Pb collisions at 17 GeV\cite{ceres_pt} are also shown for four centralities.  HIJING cacluations are also shown for Au+Au collisions.  HIJING underpredicts the observed correlations for Au+Au for all incident energies and centralities. 
\begin{figure}
\begin{center}
\includegraphics[width=15.7cm]{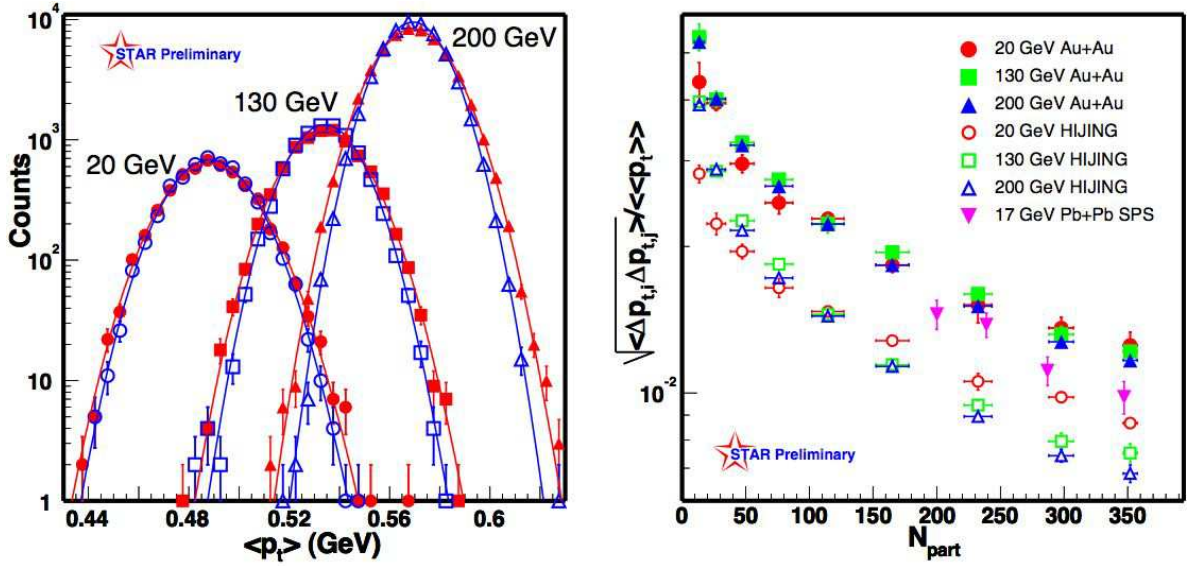}
\end{center}
\caption{\label{fig:fig2} Left panel: Histograms of the average transverse momentum per event for Au+Au at $\sqrt{s_{NN}}$ = 20 (circles), 130 (squares), and 200 GeV (triangles) for the 5\% most central collisions at each energy.  Both data (closed symbols) and mixed events (open symbols) are shown for each incident energy.  The solid lines represent gamma distribution fits. Right panel: The quantity $\sqrt {\left\langle {\Delta p_{t,i}\Delta p_{t,j}} \right\rangle }/\left\langle {\left\langle {p_t} \right\rangle } \right\rangle$ as a function of the number of participating nucleons for Au+Au at 20, 130, and 200 GeV compared with SPS results for Pb+Pb at 17 GeV\cite{ceres_pt}.  Open symbols represent HIJING calculations. } 
\end{figure}

{\em Net Charge Fluctuations.}  Another observable related to fluctuations and correlations is net charge fluctuations.  Net charge fluctuations are defined in terms of the quantity $\nu_{+-,dyn}$.  This variable is designed to quantify dynamic charge fluctuations while minimizing trivial statistical fluctuations.
Details of the net charge fluctuation analysis can be found in Refs. \cite{methods_fluctuations} and \cite{star_charge_fluc}.

In Fig. 3, results for $\nu_{+-,dyn}$ are shown for Au+Au collisions at 20, 130, and 200 GeV and for p+p collisions at 200 GeV for all charged particles with $|\eta| \le 0.5$ and $0.1 \le p_{t} \le 5$ GeV/c.  One expects that $\nu_{+-,dyn}$ will exhibit a dependence on the inverse of the multiplicity so we multiply $\nu_{+-,dyn}$ by $dN/d\eta$ in the left panel  and by the number of binary scatterings, $N_{bin}$ in the right panel.  One can see that the values for  $dN/d\eta \times \nu_{+-, dyn}$ are more negative for Au+Au at 20 GeV than those for 130 and 200 GeV while the values for $N_{bin} \times \nu_{+-, dyn}$ show little dependence on incident energy.  The scaling with $N_{part}$ holds for Au+Au collisions at all three energies as well as for p+p collisions.  The centrality dependence of the net charge fluctuations has been interpreted in terms of the onset of equilibration in central Au+Au collisions.\cite{gavin_pt_fluc}

{\em Conclusions.}  We present an overview of fluctuation and correlation measurements in STAR including the balance function, $p_{t}$ fluctuations, and net charge fluctuations.  We observe that the balance function, $B(\Delta\eta)$, narrows in central collisions consistent with trends predicted by models incorporating delayed hadronization.  We observe dynamical $p_{t}$ fluctuations at all energies at RHIC.  We quantify these fluctuations using the two particle correlation $\langle \Delta p_{t,i} \Delta p_{t,j} \rangle$, which shows little dependence on the incident energy.  Net charge fluctuations characterized by $\nu_{dyn,+-}$ scale smoothly with $N_{part}$ and have been interpreted in terms of the onset of equlibration in central Au+Au collisions.  Net charge fluctuations are similar for p+p and peripheral Au+Au collsions while central collisions may show signs of thermalization.

\begin{figure}
\includegraphics[width=15.7cm]{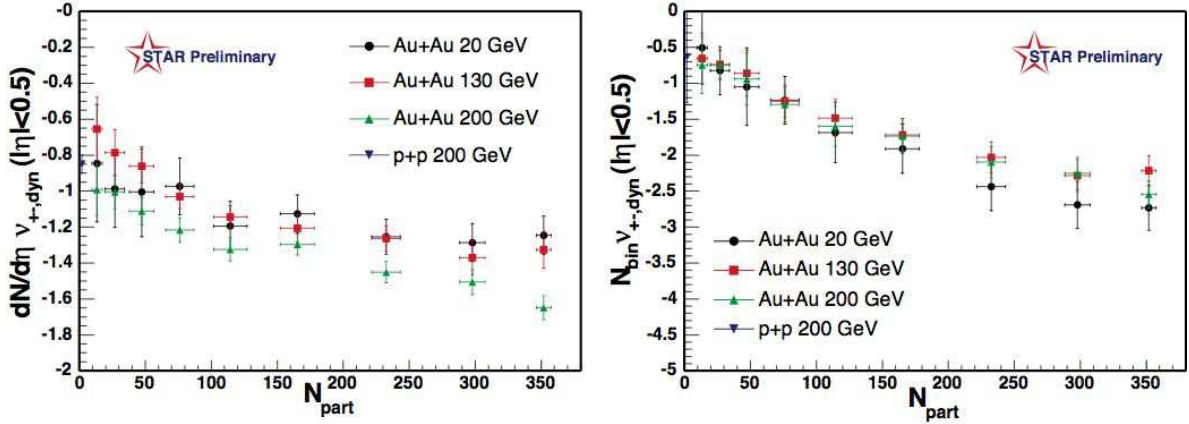}
\caption{\label{fig:fig3} Left panel: The quantity $dN/d\eta \times \nu_{+-, dyn}$ for all charged particles with $|\eta| < 0.5$ from Au+Au collisions at 20, 130, and 200 GeV and p+p collisions at 20 GeV as a function of the number of participating nucleons. Right panel: The quantity $N_{bin} \times \nu_{+-, dyn}$ for all charged particles with $|\eta| < 0.5$ from Au+Au collisions at 20, 130, and 200 GeV and p+p collisions at 20 GeV as a function of the number of participating nucleons.}
\end{figure}

\end{document}